\documentclass[english,aps,twocolumn,superscriptaddress,longbibliography]{
revtex4-1}

\pdfoutput=1
\usepackage[T1]{fontenc}
\usepackage[latin9]{inputenc}
\usepackage{color}
\usepackage{bm}
\usepackage{amsmath}
\usepackage{graphicx}

\makeatletter
\usepackage{amsmath}
\usepackage{subfigure}
\usepackage{mathtools}
\usepackage[colorlinks=true,linkcolor=MidnightBlue,urlcolor=MidnightBlue,citecolor=MidnightBlue,anchorcolor=MidnightBlue]{hyperref}\usepackage[dvipsnames]{xcolor}

\makeatother

\usepackage{babel}
\begin{document}

\title{Colloidal transport by active filaments}

\author{Raj Kumar Manna}

\affiliation{Department of Physics, Indian Institute of Technology Madras, Chennai
600036, India}

\author{P. B. Sunil Kumar}

\affiliation{Department of Physics, Indian Institute of Technology Madras, Chennai
600036, India}

\author{R. Adhikari}

\affiliation{The Institute of Mathematical Sciences-HBNI, CIT Campus, Chennai
600113, India}
\begin{abstract}
\textcolor{black}{Enhanced colloidal transport} beyond the limit imposed
by diffusion is usually achieved through external fields. Here, we
demonstrate the ballistic transport of a colloidal sphere using internal
sources of energy provided by an attached active filament. The latter
is modeled as a chain of \textcolor{black}{chemo-mechanically }active
beads connected by potentials that enforce semi-flexibility and self-avoidance.
The fluid flow produced by the active beads and the forces they mediate
are explicitly taken into account in the overdamped equations of motion
describing the colloid-filament assembly. The speed and efficiency
of transport depend on the dynamical conformational states of the
filament. We characterize these states using filament writhe as an
order parameter and identify ones yielding maxima in speed and efficiency
of transport. The transport mechanism reported here has a remarkable
resemblance to \textcolor{black}{the} flagellar propulsion of microorganisms
which suggests its utility in biomimetic systems. 
\end{abstract}
\maketitle

\section{introduction}

Diffusion is a universal but slow mechanism for particle transport
at finite temperatures. Solutions to the problem of enhancing the
rate of transport beyond the diffusion limit are found at several
scales in living systems. At the sub-cellular scale, special proteins
called molecular motors transport macromolecules super-diffusively
along microtubule tracks \citep{mitchison1996actin,vale2003molecular,wang2012active}.
At the cellular scale, molecular motors induce collective motion of
the intra-cellular fluid, a phenomenon known as cytoplasmic streaming
\citep{gutzeit1982time,theurkauf1994premature,trong2012coupling}.
At the extra-cellular scale collective motion of cilia, known as metachronal
waves, transports visco-elastic fluids along channels and provides,
alongside flagella, motility to whole organisms \citep{short2006flows,brumley2012hydrodynamic,elgeti2013emergence}.
These \emph{active }transport processes, by consuming internal sources
of energy, are able to sustain gradients in entropy, and therefore,
of particle concentration. Their ability to transport particles against
concentration gradients and over free-energy barriers has numerous
uses in biology \citep{serbus2005dynein,sawamoto2006new,gaffney2011mammalian,wilson2013high}. 

It has been notoriously difficult to synthetically replicate the active
transport solutions arrived at through many millions of years of natural
selection. Instead, enhanced particle transport in physical and chemical
contexts has largely been achieved by the input of external sources
of energy through applied fields \citep{duhr2006molecules,russel1992colloidal}.
With the coming together of physical, chemical and biological phenomena
in the domain broadly termed as biomimetics there is a fresh interest
in searching for transport mechanisms that are both active \emph{and}
synthetic. Their potential applications are numerous, including the
removal of damaged cells \citep{feynman1960there} and the targeted
delivery of drugs \citep{ishiyama2002magnetic} and microscopic devices
\citep{nelson2010microrobots}.
\begin{figure}[t]
\includegraphics[width=0.48\textwidth]{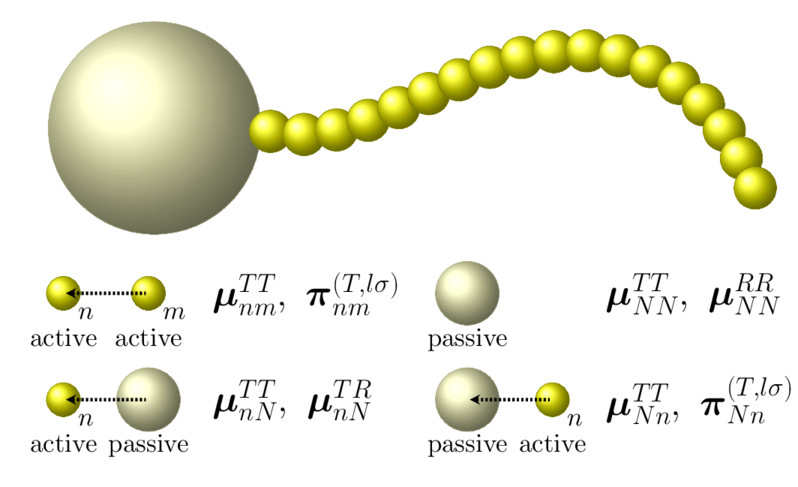}

\caption{An active filament, consisting of $N-1$ active beads (small spheres),
attached to a passive colloid (large sphere). The various hydrodynamic
interactions between the $N$ spheres, comprising the colloid-filament
assembly, are shown alongside. See text for explanation of symbols.\label{fig:active-filament}}
\end{figure}

There are several challenges in designing active transport solutions
at the microscale. First, energy has to be supplied locally to the
transport engine. In contrast to the macroscale, extracting work out
of a heat engine is unfeasible due to impossibility of maintaing heat
baths at distinct temperatures \citep{jones2004soft}. The controlled
release of chemical energy appears to be the most convenient method
of energy supply and recent successful designs borrowed from biology
in using adenosine triphosphate (ATP) as the source of energy \citep{schaller2010polar,sanchez2011cilia}.
Second, the mechanical motion that results in transport has to be
non-reciprocal, a result Purcell named as the ``scallop theorem''
\citep{purcell1977life}. This is a consequence of the dominance of
viscous forces at the microscale, where forces, to an excellent approximation,
are proportional to velocities, rather than to accelerations. Reciprocal
motion of a single degree of freedom, then, produces zero net motion.
\textcolor{black}{Therefore,} to produce net directed motion the mechanical
forcing of the fluid has to be \textcolor{black}{done} using at least
two degrees of freedom. Third, the magnitude of directed motion has
to be sufficient to overcome diffusion. Fourth, in the absence of
organizational structures like the cytoskeleton, mechanisms must be
available in the transport engine itself to navigate in three dimensional
space. Finally, for biomimetic applications the design has to be biocompatible,
avoiding hazardous chemicals or strong external fields. 

Due to the paucity of local power injection mechanisms, enhanced transport
at the micron scale is mainly achieved by applying external gradients
of electrical \citep{russel1992colloidal}, thermal \citep{duhr2006molecules}
or concentration \citep{abecassis2008boosting} fields. In one class
of mechanisms, interactions between such externally applied fields
and the particle boundary produces interfacial stresses that leads
to particle transport, collectively referred to as phoretic motion
\citep{anderson1989colloid}. Sen, Mallouk and co-workers realised
that the phoretic mechanism could be utilised by internally generated
field gradients, leading to their design of autophoretic colloidal
particles that drew on the energy released by a catalytic reaction
at the particle surface \citep{Paxton2004}. This has produced an
explosion of research in autophoresis, and more generally, in utilising
chemical energy in the solvent to transcend the diffusion limit. New
autophoretic mechanisms include bubble propulsion \citep{gao2011highly}
and redox reactions at the particle surface \citep{mallick2016autonomous}.
In another class of mechanisms, the externally applied fields act
directly on the particles. Acoustic radiation pressure from ultrasonic
standing waves has recently been used to transport micrometer size
metallic rods \citep{wang2012autonomous}. Structural chirality has
been exploited in nano-propellers that are driven by external magnetic
fields \citep{ghosh2009controlled}. Moving from rigid to flexible
objects, the beating motion of a chain of paramagnetic beads driven
by an oscillating magnetic field has been used to create propulsion
engines reminiscent of cilia and flagella \citep{dreyfus2005microscopic}.
This, by no means, is a complete survey and the reader is referred
to \citep{ebbens2010pursuit,wang2012nano,wang2013small,elgeti2015physics}
for the state of the art.

Recently, a remarkable new class of material has been created in which
internal sources of energy, provided by ATP in solution, is utilised
to generate motion \citep{sanchez2011cilia,sanchez2012spontaneous}.
The material is a mixture of microtubules, polymers that induce depletion
interactions between them, and molecular motors. The microtubules
self-assemble into filaments that beat spontaneously in the presence
of ATP, driven by the motion of the molecular motors. Such active
filaments have great potential for use in enhanced transport, as they
utilize local sources of energy, are not bound by the scallop theorem,
produce forces that are many times larger than \textcolor{black}{their
}diffusive counterparts, and are biocompatible. Navigability is yet
to be achieved using such active filaments.

Here we show that an active filament attached to a colloidal sphere
provides an active transport solution that meets all the five desiderata
listed above. We use a general \textcolor{black}{model,} that includes
hydrodynamic interactions, to describe the active filament \citep{jayaraman2012autonomous,laskar2013hydrodynamic,laskar2015brownian}.
Sustained motion is produced by a balance of forces arising from the
spontaneous activity, modeled as a distribution of stresslets along
the filament length, and the elasticity of the filament. The exchange
of momentum between fluid and filament and its local conservation
in the fluid are taken into account through an integral representation
of the fluid flow. Waveforms and beat periods obtained from this model
\citep{laskar2013hydrodynamic} are in excellent agreement with experiment
\citep{sanchez2011cilia}. Our main results are that enhanced colloidal
transport can be achieved through the active filament engine, that
speed and efficiency of the transport depend on the dynamical steady
states of the filament, that these steady states can be accurately
classified using the filament writhe as an order parameter, and finally,
that states yielding the greatest speed or efficiency can thus be
identified. We discuss how navigation can be achieved by including
a paramagnetic component in the colloid. We conclude by suggesting
several biomimetic systems where our design can be put to use.

\section{Model}

Our model for the assembly of the active filament and the colloid
consists of $n=1,\ldots,N-1$ spherical active beads of radius \textcolor{black}{$b$}
and center-of-mass coordinates $\mathbf{R}_{n}$ chained together
by potentials and a single passive sphere of radius\textcolor{black}{{}
$b_{c}\gg b$ }and center-of-mass coordinate \textbf{$\mathbf{R}_{N}$}.
The filament is clamped to the surface of the colloid through constraint
forces. A schematic is shown in Fig.(\ref{fig:active-filament}).
At low Reynolds number, Newton's equation of motion for the $N$ spheres
reduce to instantaneous balance of forces and torques,

\begin{equation}
{\bf F}_{n}+{\bf F}_{n}^{b}+\boldsymbol{\xi}_{n}^{T}={\bf 0},\,\,\,\,\,\,\,\,\,{\bf T}_{n}+{\bf T}_{n}^{b}+\boldsymbol{\xi}_{n}^{R}={\bf 0},
\end{equation}
where ${\bf F}_{n}^{b}$ and ${\bf T}_{n}^{b}$ are the net body force
and torque and $\boldsymbol{\xi}_{n}^{T}$ and $\boldsymbol{\xi}_{n}^{R}$
are the Brownian force and torque on the $n$-th sphere. ${\bf F}_{n}=\int{\bf f}\,dS_{n}$
and ${\bf T}_{n}=\int\boldsymbol{\rho}_{n}\times{\bf f}\,dS_{n}$
are the total hydrodynamic force and torque in terms of the integral
of the traction ${\bf f}={\bf n\cdot\boldsymbol{\sigma}}$, where
$\boldsymbol{\rho}_{n}$ is the vector from the center of the $n$-th
sphere to a point on its surface and $\boldsymbol{\sigma}$ is the
Cauchy stress in the fluid. 

The Cauchy stress is determined by solving the Stokes equation $\nabla\cdot\boldsymbol{\sigma}=0$
for the fluid velocity $\mathbf{v}$ together with the incompressibility
condition $\nabla\cdot\mathbf{v}=0$ and the slip boundary conditions 

\begin{equation}
{\bf v}({\bf R}_{n}+\boldsymbol{\rho}_{n})={\bf V}_{n}+\boldsymbol{\Omega}_{n}\times\boldsymbol{\rho}_{n}+{\bf v}_{n}^{\mathcal{A}}(\boldsymbol{\rho}_{n})
\end{equation}
on the surface of the $N-1$ active spheres and the usual no-slip
boundary condition on the surface of the colloid. The active slip
${\bf v}_{n}^{\mathcal{A}}(\boldsymbol{\rho}_{n})$ is conveniently
expanded in the complete orthogonal basis of irreducible tensorial
spherical harmonics, ${\bf Y}^{l}(\hat{\boldsymbol{\rho}})=(-1)^{l}\rho^{l+1}\boldsymbol{\nabla}^{(l)}\rho^{-1}$,
as 

\begin{equation}
{\bf v}_{n}^{\mathcal{A}}({\bf R}_{n}+\boldsymbol{\rho}_{n})=\sum_{l=1}^{\infty}A_{l}{\bf V}_{n}^{(l)}\cdot{\bf Y}^{(l-1)}(\hat{\boldsymbol{\rho}_{n}})
\end{equation}
where $A_{l}=1/(l-1)!(2l-3)!!$ is a normalization. The expansion
coefficients ${\bf V}_{n}^{l}$ are tensors of rank $l$, irreducible
in their last $l-1$ indices, and can thus be expressed as the sum
of three irreducible tensors ${\bf V}_{n}^{(ls)}$, ${\bf V}_{n}^{(la)}$
and ${\bf V}_{n}^{(lt)}$, of rank $l$, $l-1$ and $l-2$ respectively.
They represent the symmetric traceless, antisymmetric and pure trace
parts of $\mathbf{V}_{n}^{(l)},$ ${\bf V}^{(ls)}=\overbracket[0.7pt][2.0pt]{{\bf V}_{n}^{(l)}},$
$\mathbf{V}^{(la)}=\overbracket[0.7pt][2.0pt]{\boldsymbol{\epsilon}\cdot{\bf V}_{n}^{(l)}}$
and ${\bf V}_{n}^{(lt)}=\overbracket[0.7pt][2.0pt]{\boldsymbol{\delta}:{\bf V}_{n}^{(l)}}$.
We use the notation $\mathbf{V}_{n}^{(l\sigma)}$, with $\sigma=s,a,t$
to denote these irreducible parts, each of which are of the dimension
of velocity. It is assumed that the form of the slip and, therefore,
the values of the coefficients are specified. The velocity, $\mathbf{V}_{n}$,
and angular velocity, $\boldsymbol{\Omega}_{n}$, are to be determined,
given the slip coefficients and the external and Brownian forces and
torques.

From linearity of Stokes flow and the boundary conditions, it follows
that the hydrodynamic forces and torques must be of the form

\begin{align*}
{\bf F}_{n}= & -\boldsymbol{\gamma}_{nm}^{TT}\cdot{\bf V}_{m}-\boldsymbol{\gamma}_{nm}^{TR}\cdot\boldsymbol{\Omega}_{m}-\boldsymbol{\gamma}_{nm}^{(T,l\sigma)}\cdot{\bf V}_{m}^{(l\sigma)},\\
{\bf T}_{n}= & -\boldsymbol{\gamma}_{nm}^{RT}\cdot{\bf V}_{m}-\boldsymbol{\gamma}_{nm}^{RR}\cdot\boldsymbol{\Omega}_{m}-\boldsymbol{\gamma}_{nm}^{(R,l\sigma)}\cdot{\bf V}_{m}^{(l\sigma)},
\end{align*}
where the summation convention is implied for all repeated indices
and the slip coefficients $\mathbf{V}_{N}^{(l\sigma)}$ of the colloid
are all identically zero. The $\boldsymbol{\gamma}_{nm,}^{\alpha\beta}$,
with $\alpha,\beta=T,R$, are the usual Stokes friction tensors, yielding
drag forces proportional to $\mathbf{V}_{m}$ and $\boldsymbol{\Omega}_{m}$.
The terms proportional to $\mathbf{V}_{m}^{l\sigma}$ are active contributions
to the forces and torques due to the slip $\mathbf{v}_{m}^{\mathcal{A}}$.
The $\boldsymbol{\gamma}_{nm}^{(T,l\sigma)}$ and $\boldsymbol{\gamma}_{nm}^{(R,l\sigma)}$
are slip friction tensors associated with the $l\sigma$ mode of the
slip velocity. A method for calculating these slip friction tensors
in terms of Green's functions of Stokes flow has been provided recently
\citep{singh2016traction,singh2016crystallization} and the reader
is referred there for further details. 
\begin{figure}[!tbph]
\includegraphics[width=0.48\textwidth]{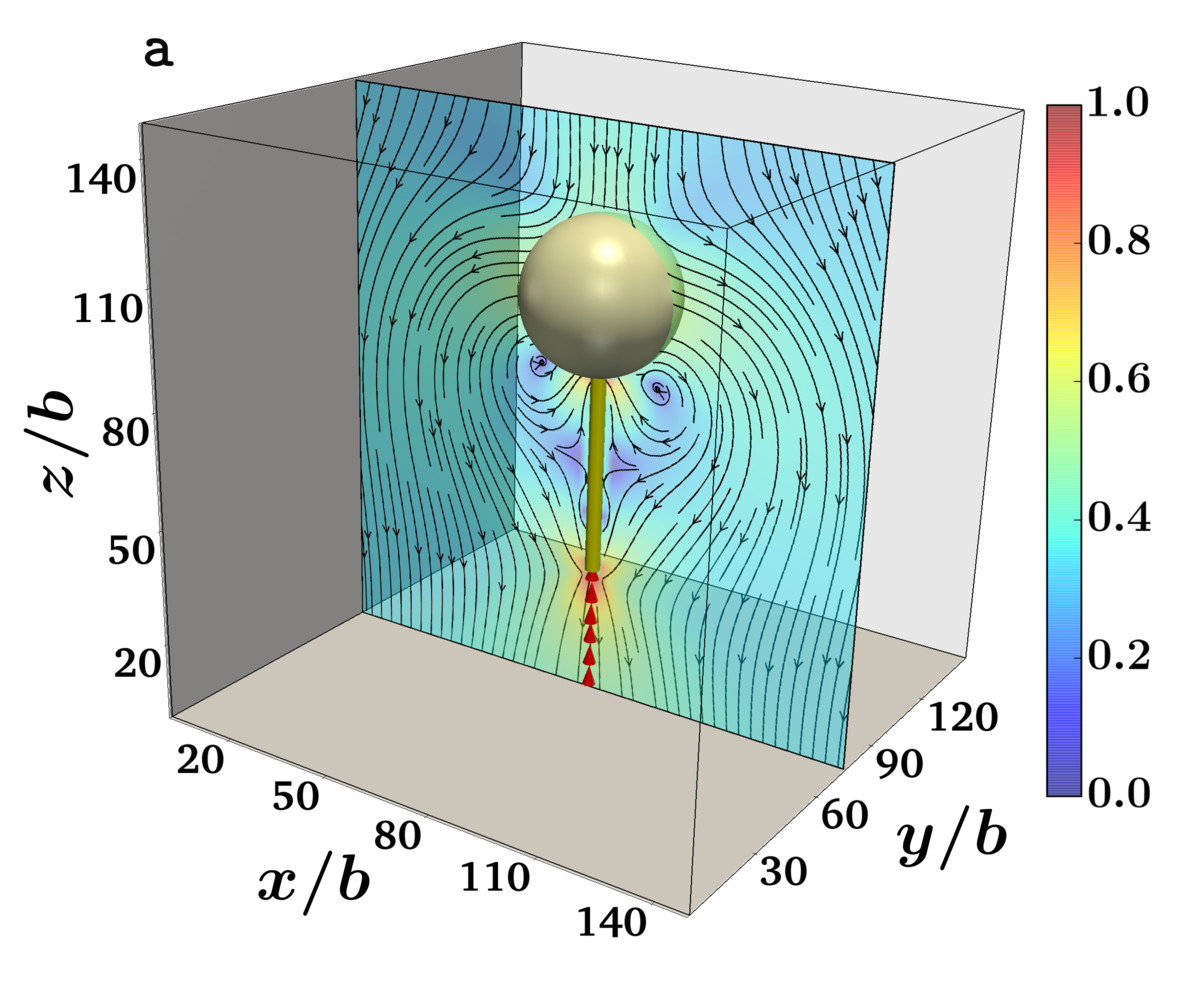}

\includegraphics[width=0.48\textwidth]{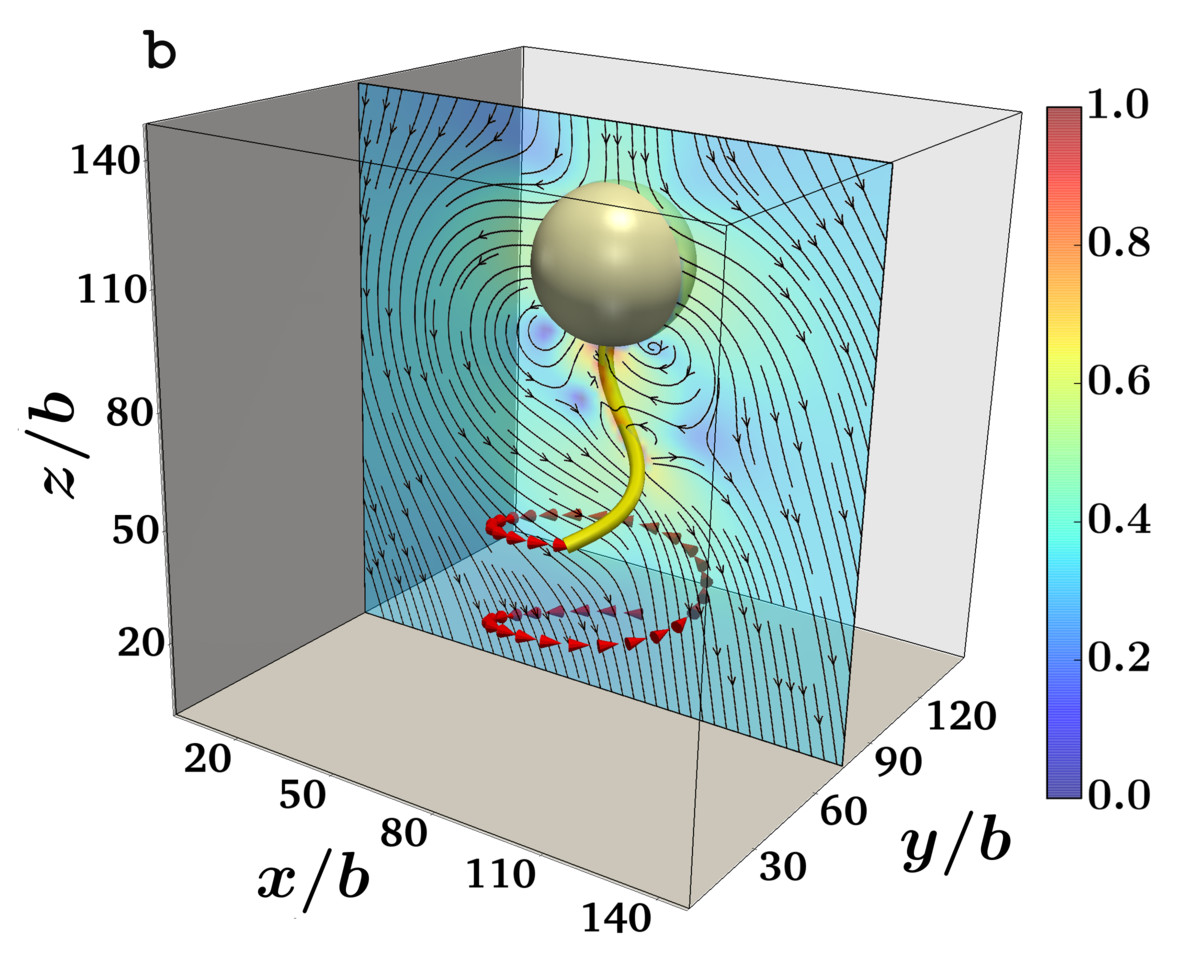}

\includegraphics[width=0.48\textwidth]{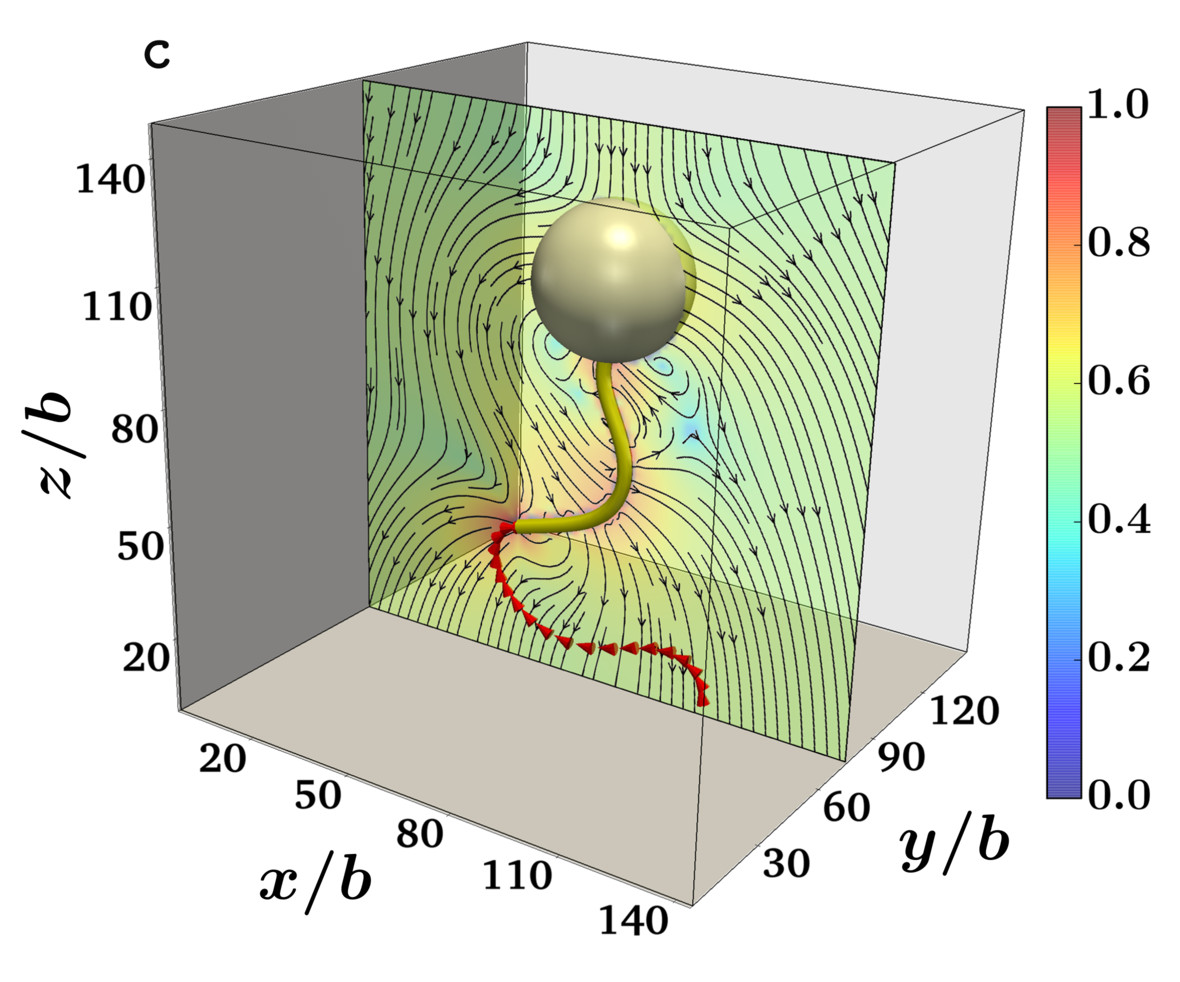}

\caption{Dynamical states of the filament-colloid assembly with varying activity
$\mathcal{A}$ showing the the \emph{linear} state at $\mathcal{A}$=10
in panel (a), the \emph{helical} state with $\mathcal{A}$=40 in panel
(b) and the \emph{planar} state at $\mathcal{A}$=80 in panel (c).
Fluid streamlines are shown in a plane passing through the equator
of the colloid, coloured by the logarithm of the magnitude of fluid
velocity normalised by its maximum. The red cones show the trajectory
of the filament terminus.\label{fig:Non-equilibrium-stationary-state}}
\end{figure}

Combining the above with Newton's equations, ignoring the Brownian
contributions and solving the resulting balances yields the following
\emph{explicit} equations for the velocity and angular velocity

\begin{align*}
\mathbf{V}_{n} & =\bm{\mu}_{nm}^{TT}\cdot\mathbf{F}_{m}^{b}+\bm{\mu}_{nm}^{TR}\cdot\mathbf{T}_{m}^{b}+\sum_{l\sigma=1s}^{\infty}\bm{\pi}_{nm}^{(T,l\sigma)}\cdot\mathbf{V}_{m}^{(l\sigma)}\\
\bm{\Omega}_{n} & =\bm{\mu}_{nm}^{RT}\cdot\mathbf{F}_{m}^{b}+\bm{\mu}_{nm}^{RR}\cdot\mathbf{T}_{m}^{b}+\sum_{l\sigma=1s}^{\infty}\bm{\pi}_{nm}^{(R,l\sigma)}\cdot\mathbf{V}_{m}^{(l\sigma)}.
\end{align*}
Here, $\mathbf{V}_{n}^{\mathcal{A}}=\mathbf{V}_{n}^{(1s)}$ is the
active translational velocity, $\bm{\Omega}_{n}^{\mathcal{A}}=b^{-1}\textbf{V}_{n}^{2a}$
is the active angular velocity, $\boldsymbol{\mu}_{nm}^{\alpha\beta}$
are the usual mobility tensors and $\bm{\pi}_{nm}^{(\alpha,l\sigma)}$
are the propulsion tensors introduced in \citep{rajesh2015}. The
relation between the mobility, slip friction and propulsion tensors
is easily verified to be 

\begin{align*}
-\boldsymbol{\pi}_{nm}^{(T,l\sigma)} & =\bm{\mu}_{nq}^{TT}\cdot\boldsymbol{\gamma}_{qm}^{(T,l\sigma)}+\bm{\mu}_{nq}^{TR}\cdot\boldsymbol{\gamma}_{qm}^{(R,l\sigma)}\\
-\boldsymbol{\pi}_{nm}^{(R,l\sigma)} & =\bm{\mu}_{nq}^{RT}\cdot\boldsymbol{\gamma}_{qm}^{(T,l\sigma)}+\bm{\mu}_{nq}^{RR}\cdot\boldsymbol{\gamma}_{qm}^{(R,l\sigma)}
\end{align*}
We evaluate all the above tensors in the pair-approximation, as is
commonly done in bead-spring models of polymers. The translational
mobility has the familiar Rotne-Prager form. Explicit forms of the
remaining tensors are provided in the Appendix. 

The above represents the equations of motion that allow for the most
general forms of surface activity. Here, we restrict ourselves to
the simplest \emph{apolar, achiral }model for slip flow, in which
the only non-zero tensorial harmonic mode corresponds to $l\sigma=2s$,
which we parametrize as\emph{ }

\begin{equation}
{\bf V}_{m}^{(2s)}=V_{0}^{(2s)}({\bf \hat{t}}_{m}{\bf \hat{t}}_{m}-\frac{1}{3}\boldsymbol{\delta})
\end{equation}
We chose the principal value $V_{0}^{(2s)}$ of this second-rank tensor
to be positive and its principal axis to be along the local tangent
${\bf \hat{t}}_{m}$ to the filament. The rich dynamical behaviour
of this minimally active filament has qualitative and quantitative
similarities with active filament systems realized experimentally
\citep{laskar2013hydrodynamic}. 

The body force $\mathbf{F}_{n}^{b}$ between the, now minimally active,
beads is obtained from the gradient of the potential $U=U^{C}+U^{E}+U^{S}$
which, in sequence, are potentials enforcing connectivity, semi-flexibility
and self-avoidance. The connectivity potential is the two body harmonic
spring potential $U^{C}({\bf R}_{m},{\bf R}_{m+1})=\frac{1}{2}k(r-b_{0})^{2}$,
where $b_{0}$ is the equilibrium bond and $r=|{\bf R}_{m}-{\bf R}_{m+1}|$.
The three-body elastic potential $U^{E}=\bar{\kappa}(1-\cos\phi)$
penalizes departures of the angle $\phi$ between consecutive bond
vectors from its equilibrium value of zero. The rigidity parameter
$\bar{\kappa}$ is related to the bending rigidity as $\kappa=b_{0}\bar{\kappa}$.
Steric effects are included through the Weeks-Chandler-Anderson potential
which vanishes if the distance between beads $r_{mn}=|{\bf R}_{m}-{\bf R}_{n}|$
exceeds $r_{\mathrm{min}}$. We assume constraint torques that result
in the vanishing rotation of the beads. Their values are obtained
from the torque balance equation with bead angular velocities set
to zero. The body force $\mathbf{F}_{N}^{b}$ and torque $\mathbf{T}_{N}^{b}$
on the colloid arise from the constraint forces that clamp the filament
to its surface. These are discussed more fully in the Appendix. 

With these specifications, the equation of motion of the active filament
is

\begin{align}
\negthickspace\!\dot{{\bf R}}_{n}= & \underbrace{\boldsymbol{\mu}_{nN}^{TT}\cdot{\bf F}_{N}^{b}+\boldsymbol{\mu}_{nN}^{TR}\cdot{\bf T}_{N}^{b}}_{\mathrm{colloid}}+\underbrace{\boldsymbol{\mu}_{nm}^{TT}\cdot{\bf F}_{m}^{b}}_{\mathrm{elasticity}}+\underbrace{\boldsymbol{\pi}_{nm}^{(T,2s)}\cdot{\bf V}_{m}^{(2s)}}_{\mathrm{activity}}\label{eq:filament-eom}
\end{align}
 In the absence of the colloid and as the activity goes to zero, the
filament equation of motion approaches the Zimm model, where $L=(N-2)b_{0}$
is the length of the filament \citep{jayaraman2012autonomous}. The
rigid body motion of the active colloid is obtained from the pair
of equations

\begin{subequations}

\begin{gather}
{\bf V}_{N}=\boldsymbol{\mu}_{NN}^{TT}\cdot{\bf F}_{N}^{b}+\boldsymbol{\mu}_{Nn}^{TT}\cdot{\bf F}_{n}^{b}+\boldsymbol{\pi}_{Nn}^{(T,2s)}\cdot{\bf V}_{n}^{(2s)}\label{eq:colloid-eom}\\
\boldsymbol{\Omega}_{N}=\underbrace{\boldsymbol{\mu}_{NN}^{RR}\cdot{\bf T}_{N}^{b}}_{\mathrm{colloid}}+\underbrace{\boldsymbol{\mu}_{Nn}^{RT}\cdot{\bf F}_{n}^{b}}_{\mathrm{elasticity}}+\underbrace{\boldsymbol{\pi}_{Nn}^{(R,2s)}\cdot{\bf V}_{n}^{(2s)}}_{\mathrm{activity}}
\end{gather}
\end{subequations}

The relative importance of activity is quantified by its ratio with
elasticity, 
\begin{equation}
\mathcal{A}=\frac{|\boldsymbol{\gamma}_{nn}^{(T,2s)}\cdot{\bf V}_{n}^{(2s)}|}{|{\bf F}_{n}^{b}|}\approx\frac{\eta b^{2}LV_{0}^{(2s)}}{\kappa}
\end{equation}
\textcolor{black}{Activity introduces a new relaxation rate $\Gamma_{s}=V_{0}^{2S}/\eta L^{3}$
in addition to rate of elastic relaxation $\Gamma_{\kappa}=\kappa/\eta L^{4}$.}
The position and orientation of the colloid changes according to the
kinematic equations $\dot{\mathbf{R}_{N}}=\mathbf{V}_{N},\quad\dot{\mathbf{p}}_{N}=\boldsymbol{\Omega}_{N}\times\mathbf{p}_{N}.$
These overdamped equations take into account the forces and torques
mediated by flow generated by the motion \emph{and} activity of the
beads and the motion of the colloid. We integrate the above set of
equations numerically to obtain the dynamics of the colloid-filament
assembly. 
\begin{figure}[b]
\includegraphics[width=0.45\textwidth]{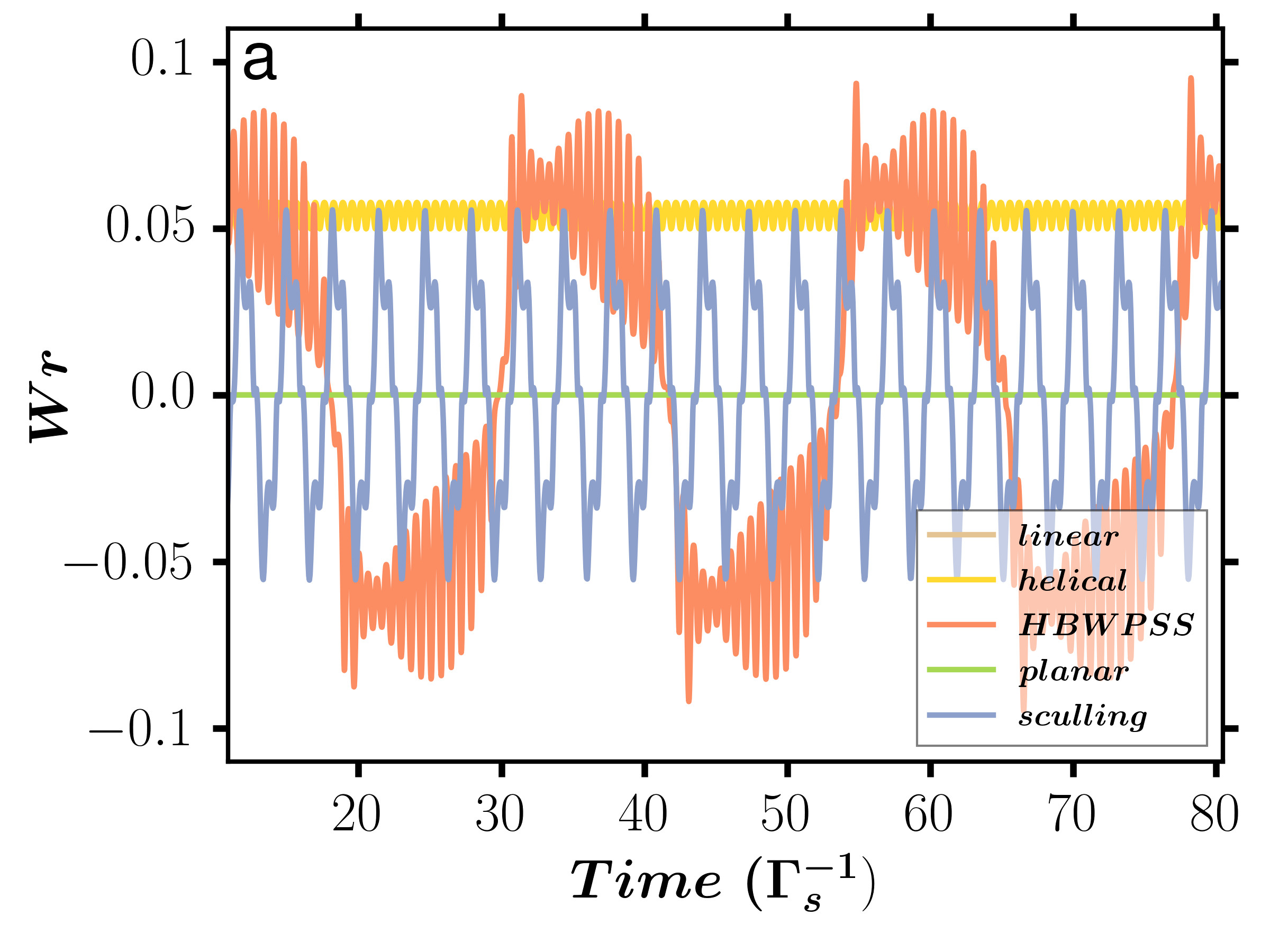}

\includegraphics[width=0.45\textwidth]{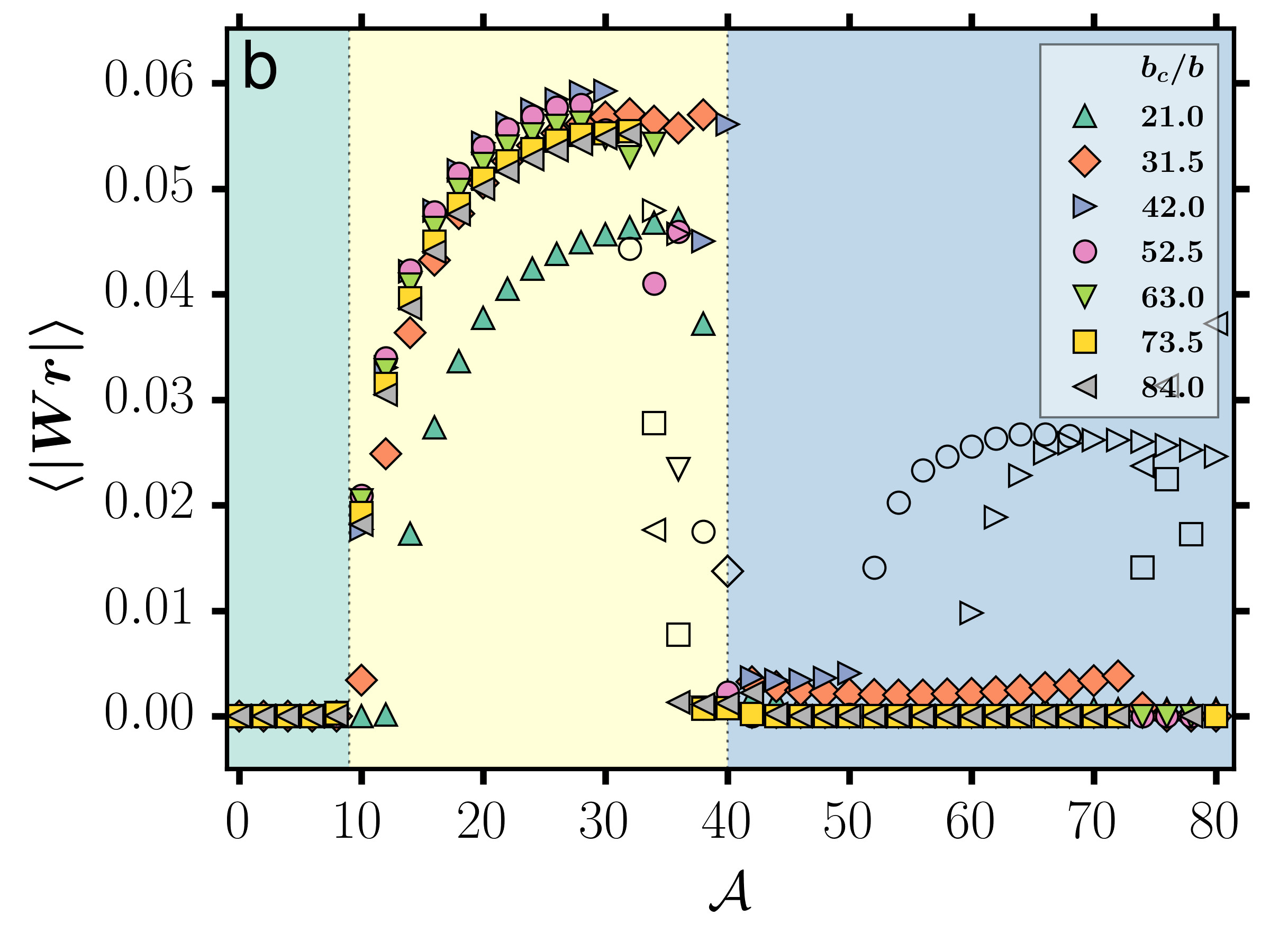} 

\caption{The time series of writhe, panel (a), and mean of its absolute value
as a function of activity $\mathcal{A}$ for $L=70b$ in panel (b).
The background colors are to mark the region of linear (green), helical
(yellow) and planar (blue) state. The empty symbols in helical and
planar region are for HBWPSS  and sculling state respectively. \label{fig:writhe}}

.
\end{figure}
\begin{figure*}[!tbph]
\includegraphics[width=0.33\textwidth]{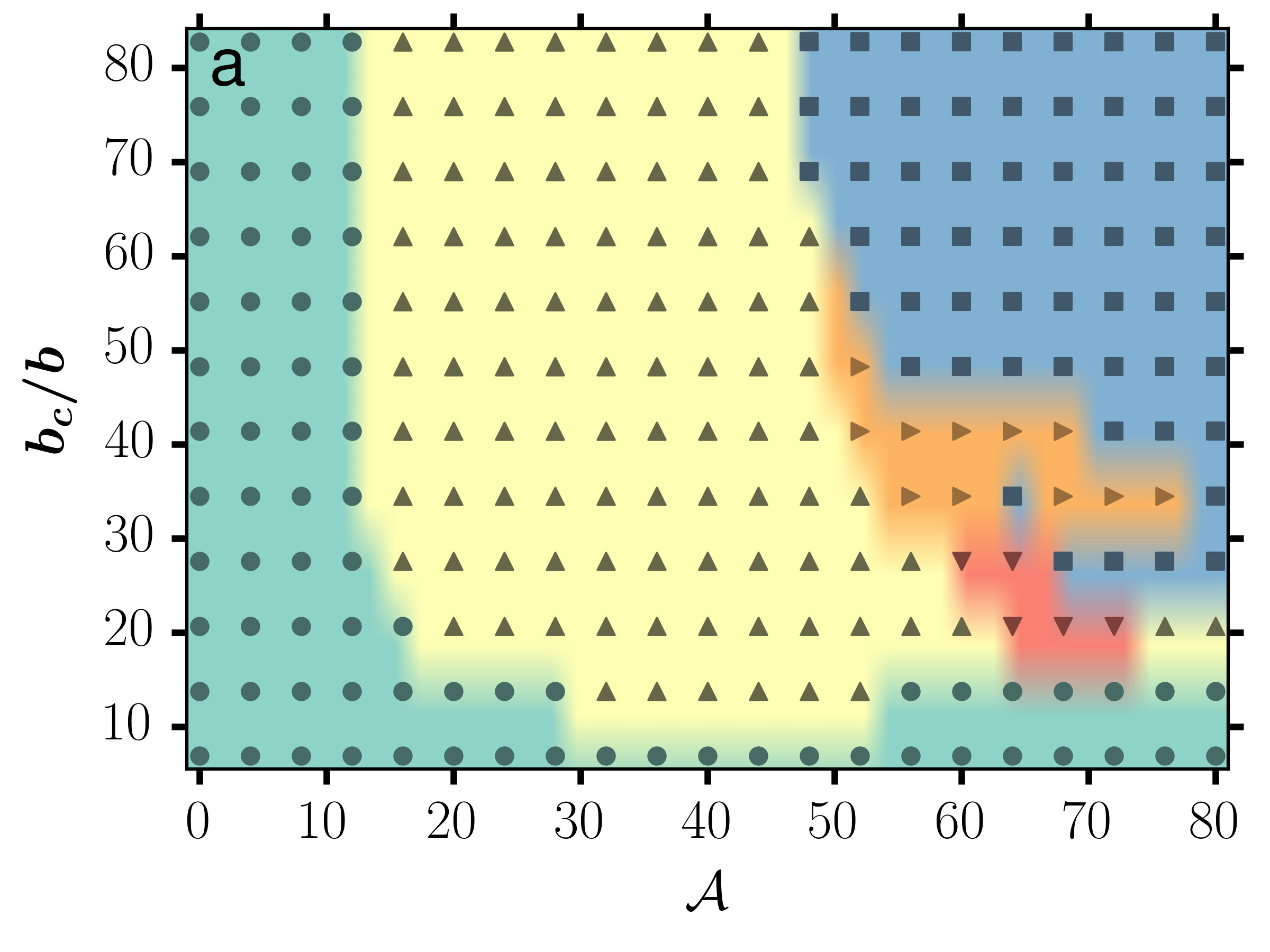}\includegraphics[width=0.33\textwidth]{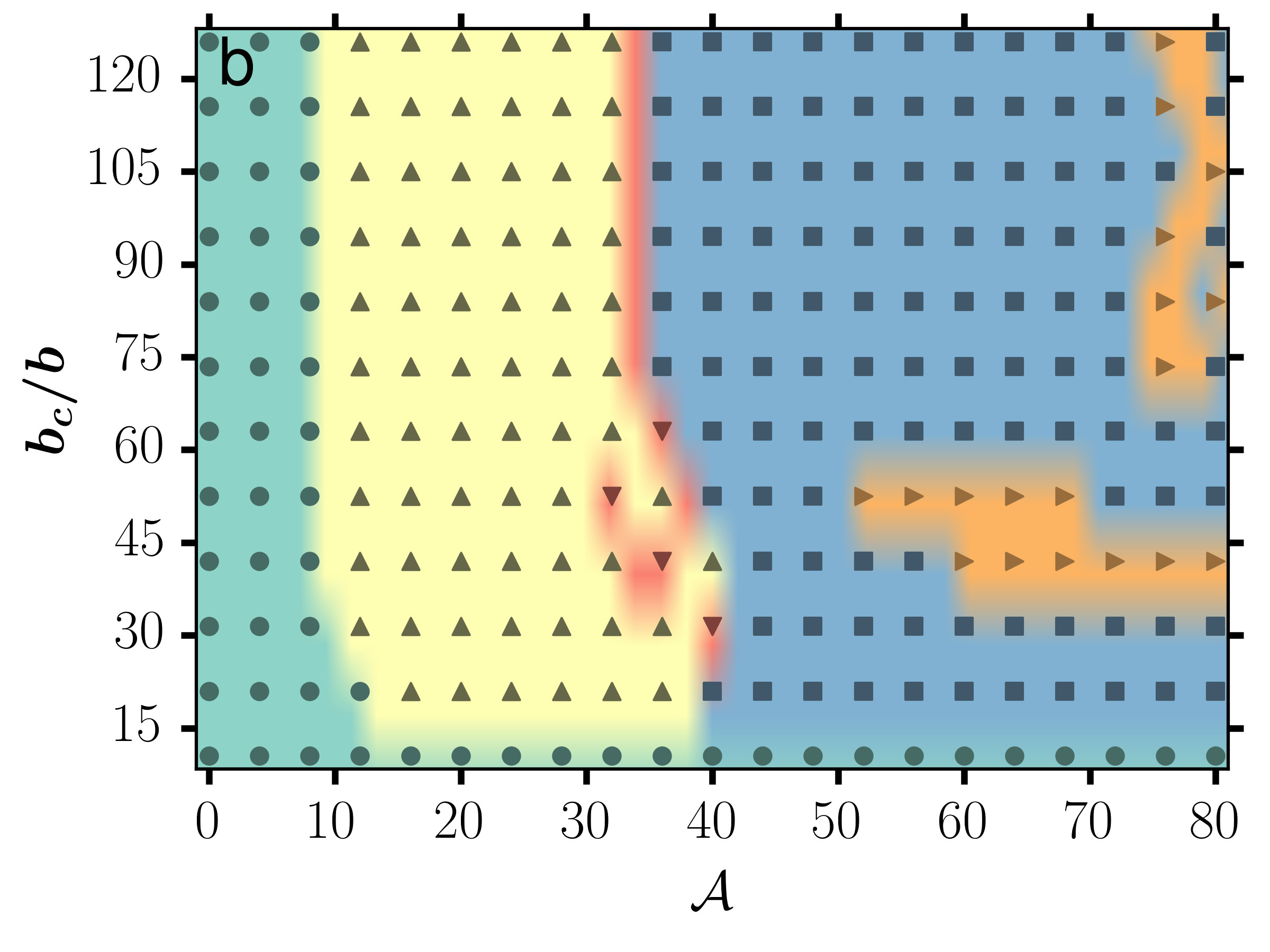}\includegraphics[width=0.33\textwidth]{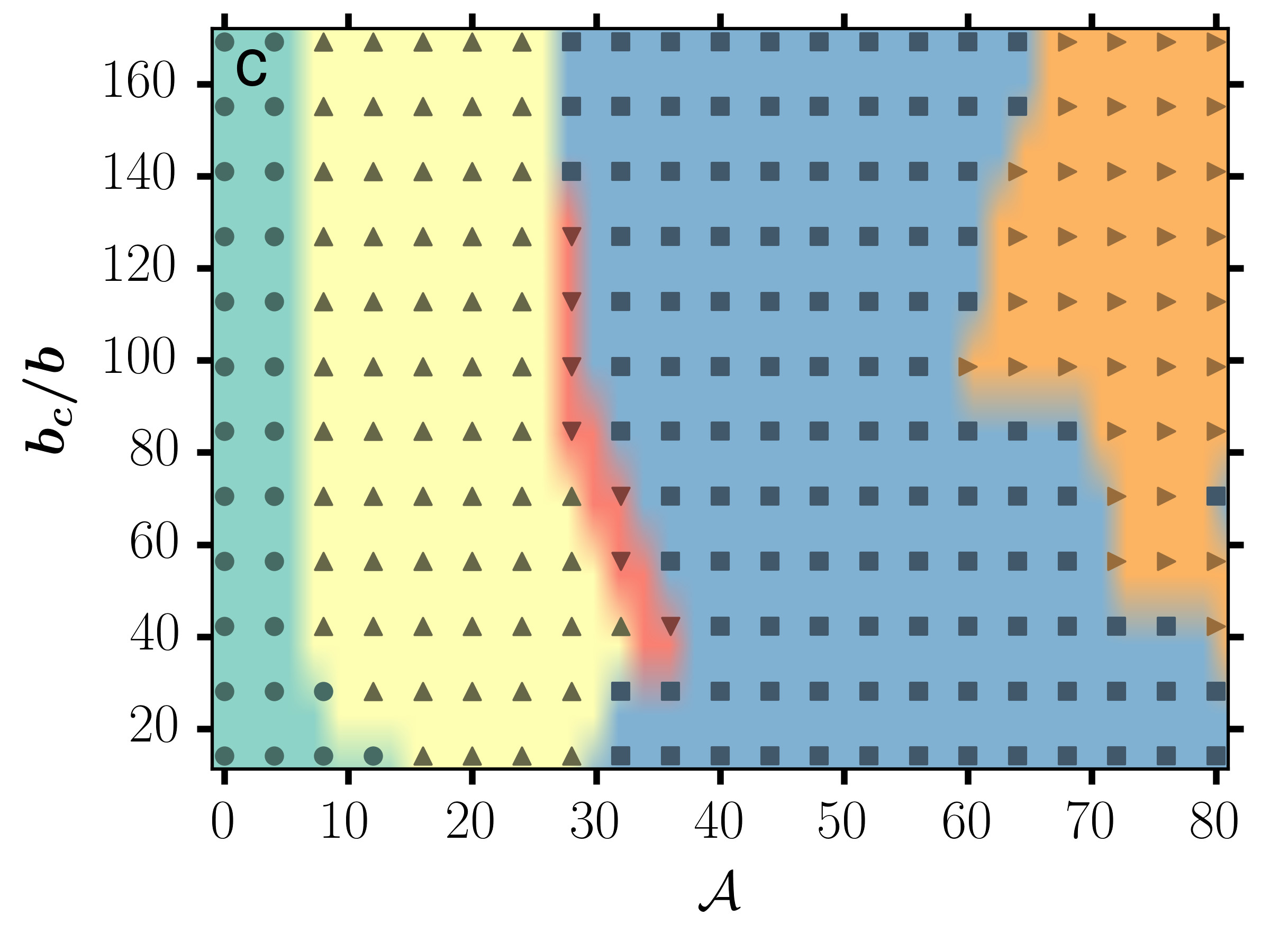}

\caption{Phase diagram of the non-equilibrium stationary states of the filament
as a function of relative size of colloid $b_{c}/b$ and dimension
less quantity $\mathcal{A}$ for three different filament lengths
$L=46b$ (a), $L=70b$ (b), $L=94b$ (c). Symbols represent the following:
linear state (circle), helical state (triangle up), helical beating
with periodic switching of sign (triangle down), planar (square) and
sculling like beating (triangle right). The colours are guides to
the eye.\label{fig:Phase-diagram-1}}
\end{figure*}

\section{Results}

We now present the results of our numerical study of the dynamics
of the colloid-filament assembly, as the ``activity number'' $\mathcal{A}$,
the relative size of the colloid \textcolor{black}{$b_{c}/b$ }and
the length $L$ of the filament are varied. 

Fig.\,(\ref{fig:Non-equilibrium-stationary-state}) shows three typical
states of motion of the assembly, with activity number increasing
from top to bottom, together with a planar section of the fluid flow
around the assembly. Panel (a) shows the simplest dynamical state,
where transport of the colloid occurs without any deformation of the
filament. Panel (b) shows a state where the filament moves rigidly
in an aplanar corkscrew-like motion, its tip tracing out a helix as
it transports the colloid. The trace of the tip is shown by the red
cones. Panel (c) shows a state in which the filament moves flexibly
in a plane, its tip tracing out an ``s'' shape while it transports
the colloid (see Movie 1 of the Supplemental Material \citep{supplementalM}).
These sequence of distinct states of motion of the filament is what
is observed when the filament is clamped at a point in an otherwise
unbounded fluid \citep{laskar2013hydrodynamic}. The principal difference
is in the values of the activity number $\mathcal{A}$ at which the
transitions take place, a difference we attribute to the modified
hydrodynamic interactions between parts of the filament arising from
the \textcolor{black}{reflection of the flow field by the surface
of the colloid. }

\textcolor{black}{These sequence of states arise from the balance
of the three kinds of forces (and torques) acting on each bead of
our filament: the conservative force from the potentials and the two
dissipative forces, one from the drag and another from the activity.
The nature of the active force, for positive $V_{0}^{(2s)}$, is such
that it produces motion }\textcolor{black}{\emph{opposite}}\textcolor{black}{{}
to the direction of the signed curvature. Therefore, as activity is
increased, states with increasing amounts of curvature appear in sequence.
Thus, in Fig. (2), we first see a linear propulsive state, then a
state in which the curvature is a fixed function of time, and finally,
states in which the curvature is a period function of time. }

A quantitative demarcation of these distinct non-equilibrium steady
states requires the introduction of an order parameter. We use the
filament writhe \citep{klenin2000computation}

\begin{equation}
Wr=\frac{1}{4\pi}\int_{C}\int_{C}d{\bf r}_{1}\times d{\bf r}_{2}\cdot\frac{{\bf r}_{1}-{\bf r}_{2}}{\vert{\bf r}_{1}-{\bf r}_{2}\vert^{3}}\label{eq:writhe-expression}
\end{equation}
 as an order parameter which can effectively distinguish these non-equilibrium
steady states. \textcolor{black}{The integrations over points $\mathbf{r}_{1}$
and $\mathbf{r}_{2}$ on the filament contour are replaced by summations
over the number of beads.} In Fig.\,(\ref{fig:writhe}), top panel,
we show the time series of writhe for the states shown in Fig.\,(\ref{fig:Non-equilibrium-stationary-state})
together with two additional states that are identified through the
order parameter. In the linear and planar states, the writhe is identically
zero. The helical state has a non-zero mean value of writhe, with
superimposed small amplitude oscillations. The two remaining states
have periodic oscillations in which the writhe averages to zero over
the cycle, but otherwise oscillates in sign. In the first of these
states, the filament rotates as in the helical state but in opposite
directions during each half of the cycle, which consists of several
periods of \emph{helical} motion. For lack of a better description,
we call this state helical beating with periodic switching of sign
(HBWPSS). The second of these states shows a motion reminiscent of
sculling, in which the filament beats in a plane that changes orientation
over the cycle (see Movie 2 of the Supplemental Material \citep{supplementalM}).
In Fig.\,(\ref{fig:writhe}), bottom panel, we show the variation
of mean of the absolute value of writhe as a function of activity
number, for different ratios of the colloid radius and filament length.
This clarifies the sequence in which the states appear. The linear
states are stable at small values of activity, being then replaced
by the helical, HBWPSS and sculling states of non-zero mean writhe
as the activity is increased. At yet larger values of activity, these
states are unstable and the planar state is generally favoured. The
non-equilibrium state diagram, thus obtained, is shown in Fig.\,(\ref{fig:Phase-diagram-1})
for three different lengths of the filament. The majority of the state
diagram is occupied by the three states shown in Fig.\,(\ref{fig:Non-equilibrium-stationary-state})
and careful parameters choices are required to locate the HBWPSS and
sculling states. 

\begin{figure}[t]
\includegraphics[width=0.45\textwidth]{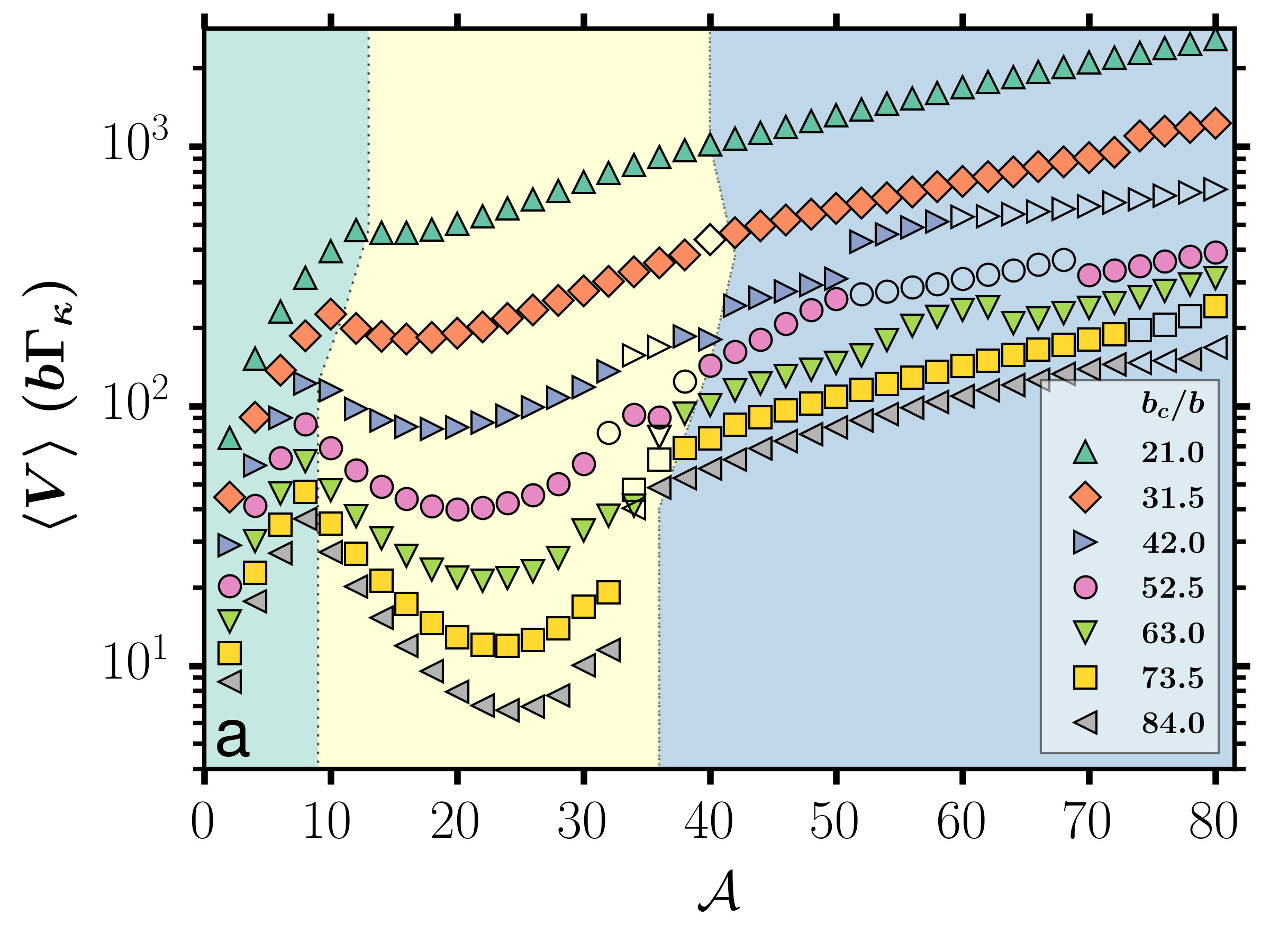}

\includegraphics[width=0.45\textwidth]{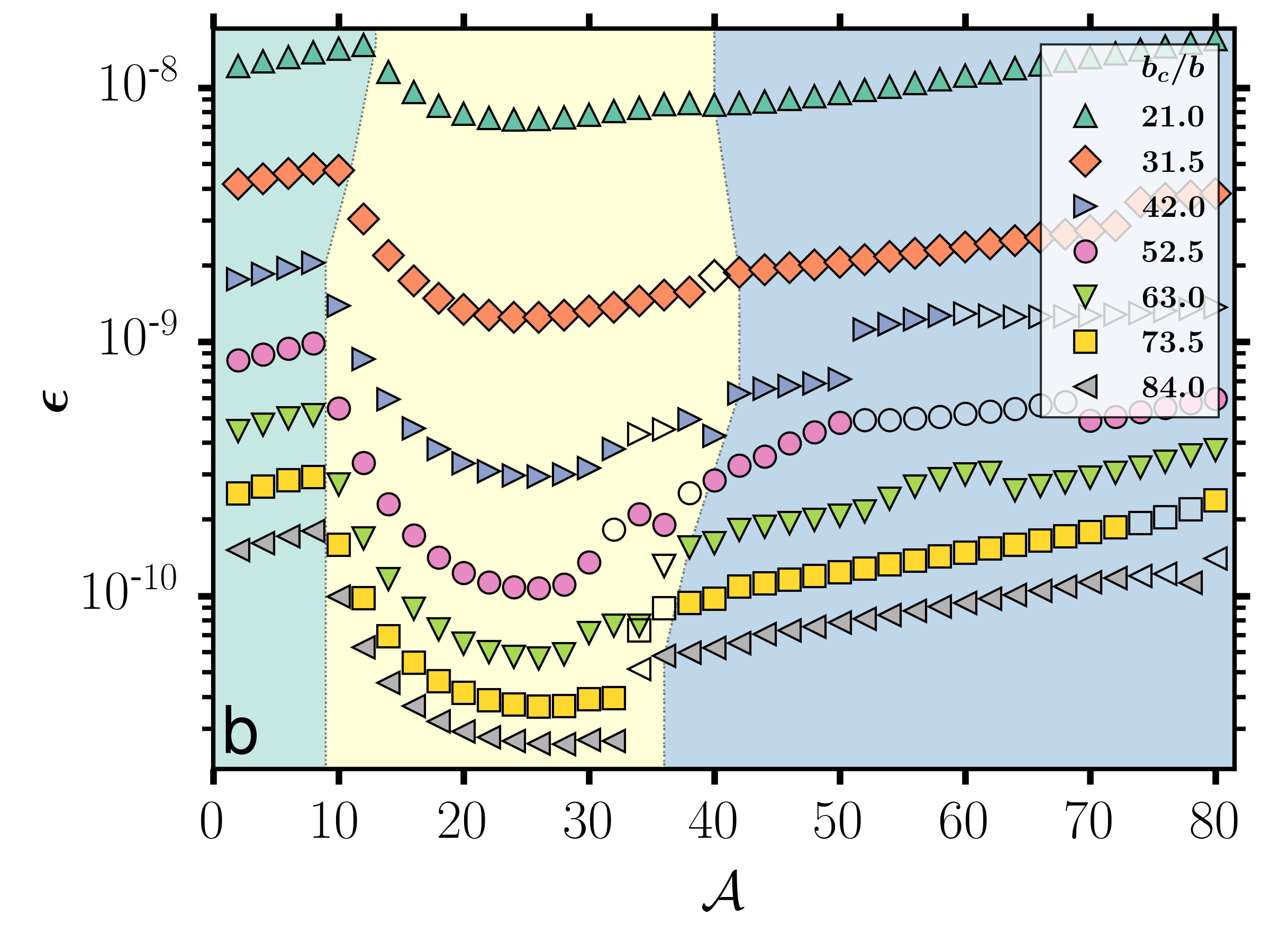}

\caption{The average translational speed a) and Efficiency of transport b)
of the colloid as a function of $\mathcal{A}$ for different colloid
radius, $b_{c}$ for filament length $L=70b$. The background colors
are to mark the region of linear (green), helical (yellow) and planar
(blue) state. The empty symbols in helical and planar region are for
HBWPSS and sculling state respectively. \label{fig:The-average-translational}}
\end{figure}

How does the speed and efficiency of colloidal transport vary across
these states? To answer this quantitatively we define, first, a measure
of efficiency, which is the ratio of the power expended in transporting
a passive colloid with velocity $\mathbf{V}$ to that expended in
the filament-colloid assembly at the same velocity, 

\begin{equation}
\epsilon=\frac{\mathbf{V}_{N}\cdot\boldsymbol{\gamma}_{NN}^{TT}\cdot\mathbf{V}_{N}}{\dot{W}}
\end{equation}
Definitions of this kind were first used by Lighthill \citep{lighthill1952}
in his study of the squirming motion of a sphere. In Fig.\,(\ref{fig:The-average-translational})
we show the variation of the speed (top panel) and efficiency (bottom
pane) of transport as a function of activity number for varying size
of the colloid. Transport speed is enhanced, more or less monotonically,
by decreasing the size of the colloid, but transport speed varies
non-monotonically with activity. States with zero mean writhe yield
greater speeds than those without, as the chemo-mechanical work done
is partially stored in the form of elastic energy in the latter class
of states, making less of it available for transport. This picture
is borne out in the variation of the efficiency $\epsilon$, where
shorter filaments in states with smaller conformational deformations
have greater efficiencies of transport. This understanding of speed
and efficiency is necessary for optimizing the design parameters of
such assemblies in possible biomimetic applications.  
\begin{figure}[h]
\includegraphics[width=0.45\textwidth]{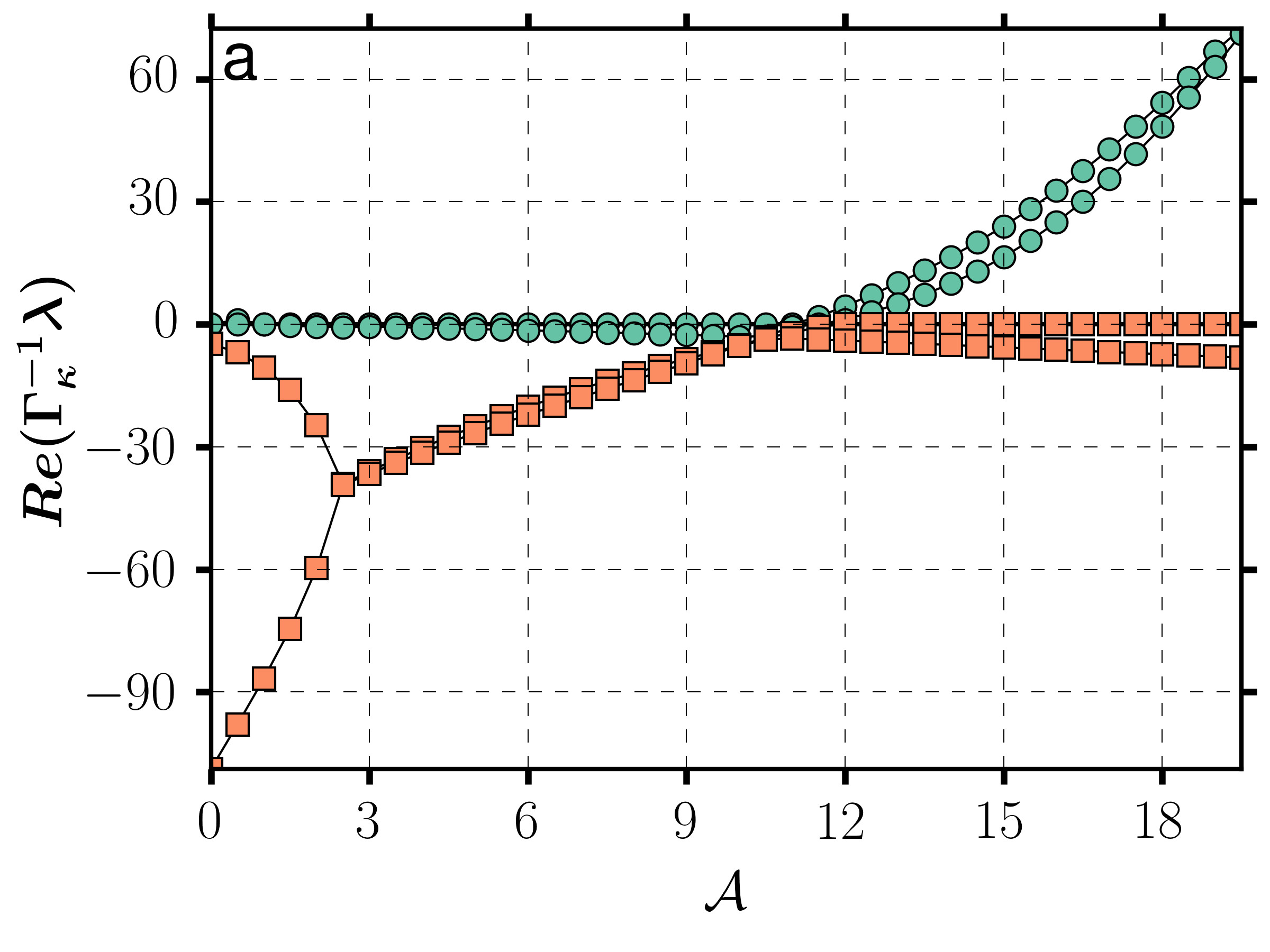}

\includegraphics[width=0.45\textwidth]{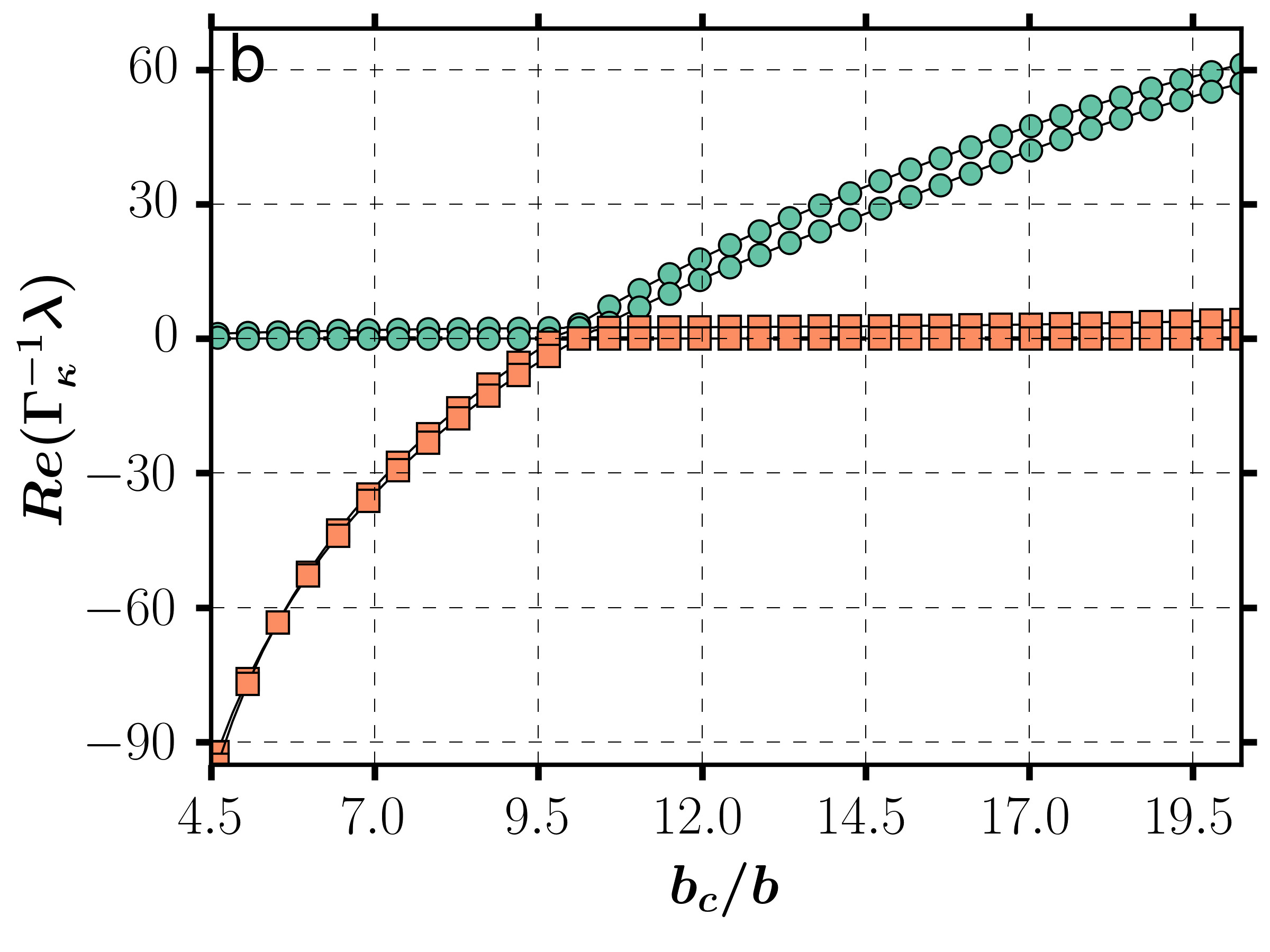}

\caption{Linear stability analysis: The largest eigenvalues of stability matrix
as function of activity number, $\mathcal{A}$ for $b_{c}=42b$ (a),
and relative colloid radius $b_{c}/b$ for $\mathcal{A}=20$ (b).
Here the length of the filament is $L=70b$. \label{fig:lsa}}
\end{figure}

The non-equilibrium stationary states are fixed points or limit cycles
of the overdamped equations of motion, Eq.\,(\ref{eq:filament-eom})
and Eq.\,(\ref{eq:colloid-eom}). It is straightforward to analyze
the linear stability of the filament as a function of activity. It
is by now well-understood that the apolar active filaments are linearly
unstable above a certain threshold value of activity \citep{jayaraman2012autonomous,laskar2013hydrodynamic,laskar2015brownian}.
This linear instability arises from the convective effect of the dipolar
flow produced by the filament. Here, we revisit the stability analysis
of \citep{laskar2013hydrodynamic}, taking into account the presence
of the colloid and the no-slip boundary conditions imposed on it.
We numerically compute the Jacobian in the linear state of the filament
and from it, obtain its largest eigenvalues, discarding the six eigenvalues
that correspond to rigid body motion. The result of this analysis
is shown in Fig.\,(\ref{fig:lsa}). In the top panel, we show the
evolution of the eigenvalues as a function of activity for filament
length $L=70b$ as we move from left to right of the middle panel
in Fig.\,(\ref{fig:Phase-diagram-1}) at $b_{c}=42b$. We see the
familiar coalescence of eigenvalues to produce a complex conjugate
pair, which then acquires a positive real part with increasing activity.
This is similar to the Hopf bifurcation seen in the case of filaments
clamped to a point in three dimensions \citep{laskar2013hydrodynamic}.
The bottom panel in Fig.\,(\ref{fig:lsa}) shows the eigenvalues
as we move from the bottom to top of the phase diagram shown in the
middle panel of Fig.\,(\ref{fig:Phase-diagram-1}) at $\mathcal{A}=20$.
Here the eigenvalues remain real but change sign from negative to
positive. This corresponds to a standard instability, rather than
a Hopf bifurcation.

\textcolor{black}{The presence of the colloid alters the value of
the activity numbers at which the bifurcations occur, but leaves unaltered
their sequence. This is because the flow reflected by the colloid
contributes only a correction to the hydrodynamic interaction mediated
by the flow produced by the beads, as an inspection of the explicit
form of this correction, provided in the Appendix, will confirm. }

\section{Discussions}

In earlier theoretical and simulation investigations on the cargo
transport by active filament, the effect of activity was introduced
by applying a local force on each bead, thus ignoring all hydrodynamic
interactions \citep{isele2016dynamics}. On the other hand, the work
presented here describes motion induced by non-local hydrodynamic
flow, resulting from active forcing, that explicitly take into account
force balance and torque balance in a three dimensional model of colloid-filament
assembly. 

We show that active filament consisting of chemo-mechanically active
apolar beads can transport a colloid to which it is clamped, in a
viscous fluid. With the over damped hydrodynamic equation of motions,
we investigate the influence of the length of the filament, the strength
of activity and the colloid radius on its motion. We identify five
different conformational states of the filament, as opposed to three
in the case of a filament clamped to a stationary point \citep{laskar2013hydrodynamic}.
It is shown that the transport speed and efficiency depend on these
dynamical states of the filament. For a given stiffness of the filament,
the speed of transport is maximum for the planar beating conformation
whereas the efficiency of transport is greatest in the case of linear
conformation. The efficiency of the transport of the colloid is found
to be in the order of $10^{-9}$, similar to that observed in the
transport of Au-Pt nano-rods in $H_{2}0_{2}$ solution \citep{paxton2005motility,wang2013small}.
The power to be delivered locally ($\sim10^{-15}$ W) and the speed
that can be achieved (several $\mu m/s$) are reasonable and are more
important parameters for application. For a given stiffness and activity
strength $V_{0}^{2s}$ the efficiency and speed are higher for lower
filament lengths ( data not shown). However, directional stability
increases with the filament length since rotational diffusivity of
an object decreases as the cube of its size. Thus we expect all regimes
of the colloid filament system to be of relevance applications. \textcolor{black}{The
propulsion efficiency is significantly lower than in bacteria, where
it varies in the range $\sim10^{-2}-10^{-4}$. This is due to the
very high efficiency of ATPase as an engine, almost $50\%$ of whose
energy is available for work. Thus, though propulsion is a small component
of the overall energy budget of an organism (Purcell has memorably
likened it to ``driving a Datsun in Saudi Arabia'' \citep{purcell1977life}),
we believe that there is room to improve the efficiency of current
active transport mechanisms. }

Though we have limited our discussion here to a system consisting
of one colloid and one filament, the equations of motion presented
here are general and can analyse more complex configurations of filaments
in an individual swimmer and a collection of swimmers. Immediate extensions
could be two filament systems like the flagella geometry of algae
Chlamydomonas, multiple filaments working synchronously in a viscous
fluid and collection of such swimmers.\textcolor{black}{{} Though the
collective dynamics of a suspension of spheres which produce constant
dipolar flows are by now well-understood, here, the far field produced
by the colloid-filament assembly is time-dependent. The collective
dynamics of such ``oscillatory dipoles'' will show new features,
such as hydrodynamic synchronization, that are absent in their time-independent
counterparts}\textcolor{blue}{{} \citep{felderhof2015stokesian,felderhof2016spinning}.}
It should be noted that the effects of wall can significantly change
the dynamics of the swimmer in the micron scale \citep{tung2015emergence}
and the states of the swimmer obtained here could be altered by such
boundary conditions. \textcolor{black}{We will present a systematic
study of these aspects in future. }

Our focus in this work has been to propose a mechanism for transport
using an active filament. Mechanisms for navigation are crucial for
applications in areas such as biomedicine. One possible navigation
mechanism is through a paramagnetic component in the colloid which
can then be controlled by an external magnetic field \citep{gauger2009fluid}.
This and other mechanisms for navigation will be presented in a future
study. 

\section*{Supplementary Material}

Movie-1\citep{supplementalM}: This movie displays the temporal behavior
of the dynamical states of the colloid-filament assembly with fluid
streamlines. 

Movie-2\citep{supplementalM}: This movie displays the dynamics of
colloid-filament assembly with filament writhe. 
\begin{acknowledgments}
We thank S. Ghose, Sachin Krishnan, A. Laskar, Rajeev Singh and Rajesh
Singh for helpful discussions. RA wishes to thank the Department of
Atomic Energy, Government of India for supporting his research. Numerical
simulations were performed on the Annapurna and HPCE clusters at The
Institute of Mathematical Sciences and IIT Madras, respectively. 
\end{acknowledgments}

\section*{APPENDIX}

\subsection*{Mobility and propulsion tensors}

The mobility tensors, are defined as 

\begin{align*}
8\pi\eta\boldsymbol{\mu}_{nm}^{TT}({\bf R}_{n},{\bf R}_{m}) & =\begin{cases}
\mathcal{F}_{n}^{0}\mathcal{F}_{m}^{0}{\bf G}({\bf R}_{n},{\bf R}_{m})\;\;\;\;\;\;\;\;\;\;\;\;\: & n\ne m\\
\frac{4}{3}b_{n}^{-1}\boldsymbol{\delta} & n=m
\end{cases}\\
8\pi\eta\boldsymbol{\mu}_{nm}^{TR}({\bf R}_{n},{\bf R}_{m}) & =\begin{cases}
\frac{1}{2}{\bf \boldsymbol{\nabla}}_{m}\times{\bf G}({\bf R}_{n},{\bf R}_{m})\;\;\;\;\;\;\;\;\;\: & n\ne m\\
{\bf 0} & n=m
\end{cases}\\
8\pi\eta\boldsymbol{\mu}_{nm}^{RT}({\bf R}_{n},{\bf R}_{m}) & =\begin{cases}
\frac{1}{2}{\bf \boldsymbol{\nabla}}_{n}\times{\bf G}({\bf R}_{n},{\bf R}_{m})\;\;\;\;\;\;\;\;\;\;\: & n\ne m\\
{\bf 0} & n=m
\end{cases}\\
8\pi\eta\boldsymbol{\mu}_{nm}^{RR}({\bf R}_{n},{\bf R}_{m}) & =\begin{cases}
\frac{1}{4}{\bf \boldsymbol{\nabla}}_{n}\times{\bf \boldsymbol{\nabla}}_{m}\times{\bf G}({\bf R}_{n},{\bf R}_{m}) & n\ne m\\
b_{n}^{-3}\boldsymbol{\delta} & n=m
\end{cases}\\
 & ,\,\,\,
\end{align*}
 where ${\bf G}$ is the Green's function for Stokes flow in un unbounded
medium,

\[
G_{ij}({\bf R}_{n},{\bf R}_{m})=\frac{\delta_{ij}}{r}+\frac{r_{i}r_{j}}{r^{3}},
\]
with ${\bf r}={\bf R}_{n}-{\bf R}_{m}$. The propulsion tensors which
relate ${\bf V}_{m}^{(l\sigma)}$, the coefficient of the traction
fields on the boundary of the $m$-th particle to the rigid body motion
are defined as 

\begin{align*}
8\pi\eta\boldsymbol{\pi}_{nm}^{(T,2s)} & =\begin{cases}
c\mathcal{F}_{n}^{0}\mathcal{F}_{m}^{1}{\bf \boldsymbol{\nabla}}_{m}{\bf G}({\bf R}_{n},{\bf R}_{m})\;\; & n\ne m\\
{\bf 0} & n=m
\end{cases}\\
8\pi\eta\boldsymbol{\pi}_{nm}^{(R,2s)} & =\begin{cases}
\tfrac{c}{2}\nabla_{n}\times{\bf \boldsymbol{\nabla}}_{m}{\bf G}({\bf R}_{n},{\bf R}_{m})\;\;\;\; & n\ne m\\
{\bf 0} & n=m
\end{cases}
\end{align*}
and $\mathcal{F}_{n}^{l}=(1+\frac{b_{n}^{2}}{4l+6}\boldsymbol{\nabla}_{n}^{2})$
is the operator.

The finite size correction to the mobility and propulsion tensors
due to the colloid are

\begin{flalign*}
 & 8\pi\eta\boldsymbol{\mu}_{nm}^{TT}({\bf R}_{n},{\bf R}_{m})=\mathcal{F}_{n}^{0}\mathcal{F}_{m}^{0}{\bf G}({\bf R}_{n},{\bf R}_{m})\\
 & +\mathcal{F}_{n}^{0}\mathcal{F}_{N}^{1}\boldsymbol{\nabla}_{N}{\bf G}({\bf R}_{n},{\bf R}_{N})\cdot(-c_{0}\mathcal{F}_{N}^{1}\mathcal{F}_{m}^{0}\boldsymbol{\nabla}_{N}\,{\bf G}({\bf R}_{N},{\bf R}_{m}))
\end{flalign*}

\begin{flalign*}
 & 8\pi\eta\boldsymbol{\pi}_{nm}^{(T,2s)}({\bf R}_{n},{\bf R}_{m})=c\mathcal{F}_{n}^{0}\mathcal{F}_{m}^{1}{\bf \boldsymbol{\nabla}}_{m}{\bf G}({\bf R}_{n},{\bf R}_{m})\\
 & +\mathcal{F}_{n}^{0}\mathcal{F}_{N}^{1}\boldsymbol{\nabla}_{N}{\bf G}({\bf R}_{n},{\bf R}_{N})\cdot(-c_{0}\mathcal{F}_{N}^{1}\mathcal{F}_{m}^{1}\,\boldsymbol{\nabla}_{N}\boldsymbol{\nabla}_{m}{\bf G}({\bf R}_{N},{\bf R}_{m}))
\end{flalign*}

\subsection*{Constraint forces}

The filament is clamped to the surface of the colloid particle. The
clamping boundary conditions are implemented by fixing the last bead
of the filament to the surface of the colloid and allowing the second
last bead to move only along the radial direction,

\begin{align*}
\dot{{\bf R}}_{N-1}={\bf V}_{N}+\boldsymbol{\Omega}_{N}\times & ({\bf R}_{N-1}-{\bf R}_{N})\\
(1-\hat{{\bf d}}\hat{{\bf d}})\cdot\Big({\bf V}+\boldsymbol{\Omega}\times{\bf d}\Big) & =(1-\hat{{\bf d}}\hat{{\bf d}})\cdot\dot{{\bf R}}_{N-2}
\end{align*}
here ${\bf d}={\bf R}_{N-2}-{\bf R}_{N}$ and $\hat{{\bf d}}={\bf d}/|{\bf d}|$.
To enforce the clamped boundary conditions on colloid, two constraint
forces ${\bf F}_{N-1}^{c}$ and ${\bf F}_{N-2}^{c}$ are applied on
the last and second last bead of the filament. These two constraint
forces are obtained by solving the above constraint equations self-consistently.
The force on the colloid particle is the negative sum of these constraint
forces. Therefore the force and torque on the colloid particle are

\begin{eqnarray*}
{\bf F}_{N} & = & -({\bf F}_{N-1}^{c}+{\bf F}_{N-2}^{c})\\
{\bf T}_{N} & = & ({\bf R}_{N-1}-{\bf R}_{N})\times{\bf F}_{N-1}^{c}+({\bf R}_{N-2}-{\bf R}_{N})\times{\bf F}_{N-2}^{c}
\end{eqnarray*}

\subsection*{Power dissipation}

The power dissipated into the fluid by the colloid-filament assembly
is

\begin{align*}
\dot{W} & =\sum_{n}-\int{\bf f}_{n}({\bf R}{}_{n}+\boldsymbol{\rho}_{n})\cdot{\bf v}({\bf R}{}_{n}+\boldsymbol{\rho}_{n})\,dS_{n}\\
 & =\sum_{n}(-{\bf V}_{n}\cdot{\bf F}_{n}^{(1s)}-\boldsymbol{\Omega}_{n}\cdot{\bf F}_{n}^{(2a)}+{\bf V}_{n}^{(l\sigma)}\cdot{\bf F}_{n}^{(l\sigma)})\\
 & =\sum_{n}({\bf V}_{n}\cdot{\bf F}_{n}^{b}+\boldsymbol{\Omega}_{n}\cdot{\bf T}_{n}^{b})+\sum_{n,m}{\bf V}_{n}^{(l\sigma)}\cdot\boldsymbol{\gamma}_{nm}^{(l\sigma,l'\sigma')}\cdot{\bf V}_{m}^{(l'\sigma')}
\end{align*}

The equations of motion of the filament beads and colloid particle
are integrated by numerically. We use spring constant $k=5.0$, equilibrium
bond length $b_{0}=2b$, rigidity parameter $\bar{\kappa}=1.6$. We
chose the number of beads $N$ in the range $24$ to $48$ and $V_{0}^{(2s)}$
in the range $0$ to $2.72.$

\bibliography{colloid-transport}

\end{document}